\documentclass[aps,pra,twocolumn]{revtex4}
\usepackage[usenames]{color}
\newcommand{\comment}[1]{}
\usepackage{graphicx}
\usepackage{epsfig}
\usepackage{amsmath}

\begin{document}
\title{Vortex Synchronization in 
Bose-Einstein Condensates: A Time-Dependent Gross-Pitaevskii Equation Approach}
\author{Ryan Barnett,$^{1,2}$ Edward Chen,$^1$ and Gil Refael$^1$}
\affiliation{$^1$Department of Physics, California Institute
of Technology, MC 114-36, Pasadena, California 91125, USA}
\affiliation{$^2$Joint Quantum Institute and Condensed Matter Theory Center,
Department of Physics, University of Maryland, College Park,
Maryland, 20742, USA}
\date{\today}
\begin{abstract}
  In this work we consider vortex lattices in rotating Bose-Einstein
  Condensates composed of two species of bosons having different
  masses.  Previously \cite{barnett08} it was claimed that the
  vortices of the two species form bound pairs and the two vortex
  lattices lock. Remarkably, the two condensates and the external
  drive all rotate at different speeds due to the disparity of the
  masses of the constituent bosons.  In this paper we study the system
  by solving the full two-component Gross-Pitaevskii equations
  numerically.  Using this approach we verify the stability of the
  putative locked state which is found to exist within a disk centered
  on the axis of rotation and which depends on the mass ratio of the
  two bosons.  We also show that an analytic estimate of this locking
  radius based on a two-body force calculation agrees well
  with the numerical results.
\end{abstract}
\maketitle

\section{Introduction}

One of the most striking manifestations of the quantum-mechanical
nature of superfluids under rotation is the formation of vortices
\cite{onsager49,feynman55}.  Perhaps the most natural arena to
controllably study the physics of vortices are Bose-Einstein
Condensates (BECs) of alkali atoms \cite{matthews99, madison00,
  abo-shaeer01}.  For the simplest case where the condensate is
composed of a single type of atom without spin degrees of freedom, a
triangular lattice is formed \cite{tkachenko66,donnelly91}.  On the
other hand, for multicomponent systems (composed of mixtures of atoms
or spinor condensates, for instance), the order parameter has additional
degrees of freedom resulting in more complex vortex lattice structures
(see, for instance, 
\cite{mueller02, kasamatsu03, mueller04, barnett08b, cooper08}).

For the classic problem of an ideal fluid in a container rotating at
rate ${\bf \Omega}$, the steady-state local velocity is ${\bf v}={\bf
  \Omega}\times {\bf r}$ where ${\bf r}$ is the distance from axis of
rotation.  Since this velocity has everywhere a nonvanishing curl
($|\nabla \times {\bf v}|=2\Omega$), it is not a permissible flow for
a superfluid which is supposed to be inherently irrotational (${\bf
  v}=\nabla\theta$ with $\theta$ the phase of the SF order parameter).
However, superfluids are well-known to mimic the classical rigid-body
rotation \emph{on average} by forming a vortex lattice where the
density of these vortices is given by the Feynman relation
\cite{feynman55,donnelly91}
\begin{equation}
\label{Eq:Feyn}
\rho_v = \frac{m\Omega}{\pi\hbar}
\end{equation}
where $m$ is the mass of the constituent bosons and $\Omega$ is the
rate the superfluid is rotating which is equal to the rotational
rate of the walls of the container.

In a previous work \cite{barnett08}, it was considered how the
situation described above generalizes to the problem of two-component BECs
composed of atoms having different masses.  More specifically,
Eq.~(\ref{Eq:Feyn}) naturally generalizes for two-component systems to
\begin{equation}
\label{Eq:Feynmulti}
\rho_v^1 = \frac{m_1\Omega_1}{\pi\hbar} \;\; ; 
\;\; \rho_v^2 = \frac{m_2\Omega_2}{\pi\hbar}
\end{equation}
where $m_1$ and $m_2$ are the masses of the bosons in the two
constituent condensates and $\Omega_1$ and $\Omega_2$ are angular
rates at which the two superfluids are rotating.  For the case where
there is a negative interspecies scattering length, the attraction
between species will lead to an attractive interaction between
vortices of the two species.  When this interaction is sufficiently
large one has the situation where the vortices form bound pairs,
forcing the densities of the two vortex lattices to be the same:
$\rho_v^1 \approx \rho_v^2$.  For this case,
Eqns.~(\ref{Eq:Feynmulti}) imply that the two superfluids (taking
without loss of generality $m_1>m_2$) will rotate at different speeds
$\Omega_1<\Omega_2$.  This counterintuitive state results from the
quantum mechanical nature of the superfluid and has no analog in the
classical fluid case.

In \cite{barnett08} the existence of this state was argued by making
an ansatz for the short-ranged interspecies interaction and performing
a two-body force calculation using it.  This gave a quantitative
prediction for the distance from the center of the condensate at which
the vortex pairs become unbound, resulting from the growth of the
Magnus force, which is referred to as the locking radius.  The goal of
the current paper is to test these arguments by numerical integration
of the full two-component Gross-Pitaevskii equation.  We will verify
that such locked states are stable for a range of parameters.
Furthermore, we will see that the analytic prediction for the locking
radius agrees well with the numerical results.

This paper is organized as follows.  First in Sec.~\ref{Sec:defs} we
provide definitions and set the notation for the treatment of the
Gross Pitaevskii equation.  In Sec.~\ref{Sec:two-body} we summarize
the derivation of the locking radius previously given in
\cite{barnett08} which is based on a two-body force calculation.  Then in
Sec.~\ref{Sec:numericalmethods} we describe the split-operator
technique utilized to propagate the Gross Pitaevskii equations in
imaginary time.  The main results of the paper are presented in
Sec.~\ref{Sec:results}.  Here we provide the vortex lattice structures
determined numerically, and compare them with the estimate for the locking
radius.  Finally, in Sec.~\ref{Sec:exp} we provide a discussion of
potential experiments to realize this effect and then conclude.

\section{Background}

\subsection{Two-component Gross-Pitaevskii Equations}
\label{Sec:defs}

Our analysis starts with the two-component
Gross Pitaevskii energy functional in the frame of reference rotating
at angular rate $\Omega_d$.  This is given by $E=E_1+E_2+E_{12}$ where
\begin{align}
E_{1} &= \int d^2r \left( \frac{\hbar^2}{2m_1} |\nabla \psi_1|^2
+V_{1}n_1
+\frac{1}{2} g_1 n_1^2 -  
\Omega_d \psi_1^* L_z \psi_1 \right), \\
E_{2} &= \int d^2r \left( \frac{\hbar^2}{2m_2} |\nabla \psi_2|^2
+V_2 n_2+
\frac{1}{2} g_2 n_2^2 -  
\Omega_d \psi_2^* L_z \psi_2 \right),  
\label{GGPP}
\end{align}
and
\begin{align}
\label{Eq:interspecies}
E_{12} &= g_{12} \int d^2 r \;  n_{1}({\bf r}) n_{2}({\bf r}).
\end{align}
In these equations, $V_1$ and $V_2$ are the confining potential of the
BECs which we will take to be harmonic.  The intraspecies and
interspecies scattering strengths are defined as $g_{1,2}$ and
$g_{12}$ respectively.  The angular momentum operator, as usual, is
defined as $L_z=xp_y - y p_x$ where $p_{x,y}\equiv -i
\hbar \partial_{x,y}$.

Varying this energy with respect to $\psi_1$ and $\psi_2$ and
introducing the chemical potentials $\mu_1$ and $\mu_2$
to enforce particle
number conservation gives the two-component Gross-Pitaevskii equations
\begin{align}
\mu_1 \psi_1 = 
- \frac{\hbar^2}{2m_1} \nabla^2 \psi_1+V_1\psi_1 + g_1 n_1 \psi_1 + g_{12}n_2 \psi_1 
-   \Omega_d  L_z \psi_1  \\
\mu_2 \psi_2 = 
- \frac{\hbar^2}{2m_1} \nabla^2 \psi_2+V_2\psi_2 + g_2 n_2 \psi_2 + g_{12} n_1 \psi_2 
-   \Omega_d  L_z \psi_2.
\end{align}
In Sec.~\ref{Sec:numericalmethods}, we will describe how these coupled
equations are solved numerically to find minima of the energy $E$.

\subsection{Inter-species vortex attraction and locking}
\label{Sec:two-body}

\begin{figure}
\includegraphics[width=3.4in]{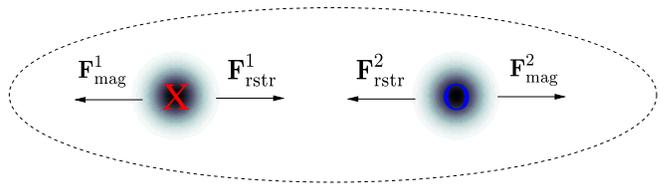}
\caption{A bound pair of vortices occurring in the locked state
  composed of a mixture of BECs with $m_1>m_2$.  The vortex in the
  heavier species is denoted with an `x' while that in the lighter
  species is denoted with an `o'. The center of the condensate is
  taken to be to the left of this bound pair.  The attractive
  short-ranged interspecies force ${\bf F}_{rstr}^{1,2}$ serves to
  bind the vortex pairs together.  This is counterbalanced by the
  Magnus force ${\bf F}_{\rm mag}^{1,2}$ which increases from the
  center of the condensate.  Note that the healing length of the
  superfluid is larger than the sphere representing vortices in this
  figure}
\label{Fig:forces}
\end{figure}

In this section, for completeness, we provide an estimate of the
locking radius of the vortex-bound state based on a two-vortex
calculation.  This calculation is presented in more detail in
\cite{barnett08}.  We first consider the state for the case where the
interspecies interaction is large and all of the vortices of one
species are bound with that of the other due to the strong
short-ranged attractive force.  A bound pair of vortices is depicted
in Fig.~\ref{Fig:forces}.  The vortex binding causes the two
superfluids to rotate at different rates (because of the different
masses and the Feynman relation), creating a Magnus force which tries
to rip the bound pair apart.  The Magnus force is balanced by the
short-ranged interspecies vortex interaction resulting from the
overlap of the vortex cores.  However, since the Magnus force grows
linearly with the distance from the center of the condensate, it will
eventually overcome the short-ranged interspecies attraction. The
point at which this occurs we refer to as the locking radius.

We will now put the previous arguments on more quantitative footing.
A vortex sitting at rest in a superfluid flowing at velocity ${\bf v}$
will experience a force perpendicular to the flow
\begin{equation}
{\bf F}_{\rm mag} = 2\pi \hbar n_0 {\bf v} \times \hat{ \bf \kappa},
\end{equation}
the so-called Magnus force \cite{donnelly91}, where $\hat{\bf \kappa}$
is a unit vector centered on the vortex pointing out the plane.
Assuming the system is composed of entirely locked vortex lattices,
we have for the vortex densities $\rho_v^{(1)} = \rho_v^{(2)}$.  This,
via the Feynman relations, gives 
\begin{equation}
m_1 \Omega_1 = m_2 \Omega_2
\end{equation}
where $\Omega_1$ and $\Omega_2$ are the angular rotational rates of
the two superfluids.  Since $m_1 > m_2$ the superfluids will rotate at
different rates which will lead to the Magnus forces
pulling the bound pair apart.

The short-ranged interspecies interaction arising from 
$E_{12}$ defined in Eq.~(\ref{Eq:interspecies})  will 
counteract the Magnus force.  
We refer to this as the ``restoring force''.  In order
to obtain an analytic expression for this interaction, we take a
Gaussian ansatz for the density profile about a vortex.  Specifically,
for a vortex in species $\alpha$ centered at ${\bf r}_0$, we take
\begin{equation}
\label{Eq:gaussden}
n_{\alpha}({\bf r})=n_0^{\alpha}(1-e^{-|{\bf r}-{\bf r}_0|^2/\lambda_\alpha^2})
\end{equation}
where $\lambda_{\alpha}$ is on the order of the superfluid coherence
length \cite{barnett08}.  We
take 
\begin{equation}
\label{Eq:lambda}
\lambda_\alpha = 1.781 \xi_\alpha
\end{equation} 
for the value of this length parameter.  This is a slight
modification of the analysis in \cite{barnett08} where the simpler
case of  $\lambda_\alpha = \xi_\alpha$ was taken.  We choose the
the nonuniversal constant in Eq.~(\ref{Eq:lambda}) so that 
the ansatz in Eq.~(\ref{Eq:gaussden}) provides a better fit 
to the density surrounding a vortex.  For a more detailed
discussion of this, see the Appendix \ref{Sec:A1}.

Inserting this ansatz for two vortices separated by distance $d$
into Eq.~(\ref{Eq:interspecies}) we find
\begin{equation}
E_{12}=g_{12} n_0^1 n_0^2 \pi
\frac{\xi_1^2 \xi_2^2}
{\lambda_1^2+\lambda_2^2}
e^{-d^2/(\lambda_1^2+ \lambda_2^2)}
\label{Eq:E12}
\end{equation}
where we have dropped terms which do not depend on the vortex
separation.  The interspecies force immediately follows from the
derivative of this interaction energy and is
\begin{equation}
\label{Eq:isforce}
{\bf F}_{\rm{rstr}}^{\alpha}=-2\pi |g_{12}| n_0^{1}
n_{0}^2\frac{\lambda_1^2\lambda_2^2}{(\lambda_1^2+\lambda_2^2)^2}
e^{-d^2/(\lambda_1^2+ \lambda_2^2)} {\bf d}.
\end{equation}
Balancing the forces on each vortex in the frame of reference
rotating at the drive frequency, we have $F_{\rm mag}^1= F_{\rm
  rstr}^1$ and $F_{\rm mag}^2= F_{\rm rstr}^2$ , as is illustrated in
Fig.~\ref{Fig:forces}.  Since the restoring force acting on either
species has the same magnitude we have that $F_{\rm mag}^1= F_{\rm
  mag}^2$.  The Magnus force on species $\alpha$ in the frame rotating
with the vortex lattice at frequency $\Omega_v$ is given by
\begin{equation}
F_{\rm mag} ^{\alpha} = 2\pi \hbar n_o^\alpha |\Omega_\alpha - \Omega_v| r
\end{equation}
which grows linearly with the distance from the center of the condensate $r$.
Also, note that the restoring force will not depend on the position in the
condensate.  A bound vortex pair will become unstable when the Magnus force
is equal to the maximum possible value of the restoring force.  This can
be worked out  to be
\begin{align}
\notag
r_c&=\sqrt{\frac{1}{2e}} \frac{|g_{12}|}{\hbar \Omega_v}
\frac{m_1 n_0^{2}+m_2 n_0^1}{m_1-m_2} \; \frac{\lambda_1^2\lambda_2^2}{(\lambda_1^2+\lambda_2^2)^{3/2}}
\\
&=
0.7638 \; \frac{|g_{12}|}{\hbar \Omega_v}
\frac{m_1 n_0^{2}+m_2 n_0^1}{m_1-m_2} \; 
\frac{\xi_1^2\xi_2^2}{(\xi_1^2+\xi_2^2)^{3/2}}.
\label{Eq:rc}
\end{align}
A more detailed derivation of
this expression can be found in Ref.~\cite{barnett08}.

\section{Numerical methods}
\label{Sec:numericalmethods}

Since there are only rare occasions when the time-dependent
Gross-Pitaevskii equation permits analytic solutions, numerical
simulation is often the method of choice for theoretically studying
Bose-Einstein condensates (for a recent account numerical solution of
the GPE, see Ref.~\cite{succi05}).  In this section, for simplicity, we
will only consider the single-component case, noting that the
generalization to the two-component case is straightforward.  To this
end, the equation we wish to solve is
\begin{equation}
\label{Eq:imtime}
\hbar \frac{\partial \psi}{\partial \tau} = H \psi
\end{equation}
which describes the evolution of $\psi$ in imaginary time, $\tau=it$.
Under long enough evolution $\psi$ will relax to the ground state of
the Gross-Pitaevskii energy functional.  In the above equation $H$ is
given by
\begin{equation}
\label{Eq:H}
H=-\frac{\hbar^2}{2m} \nabla^2 + V_{\rm trap} + g |\psi|^2 - \Omega L_z.
\end{equation}

We use a split-operator method to evolve the order parameter $\psi$ as
in Eq.~(\ref{Eq:imtime}).  The idea behind the split-operator method
is to approximate the evolution operator through imaginary time
interval $\Delta \tau$, $U(\Delta \tau)=e^{-H \Delta \tau}$, by a
product of terms which are easily diagonalizable.  
Neglecting for the moment the rotational term in Eq.~(\ref{Eq:H}), $H$
can be written as the sum of two terms, $H=T+V$, where 
$
T=-\frac{\hbar^2}{2m} \nabla^2
$
and 
$
V=V_{\rm trap} + g |\psi|^2
$.
These terms are easily diagonalized
in momentum and position space respectively.  The wave function
$\psi$ can then be advanced in time by $\Delta \tau$ by 
\begin{align}
\label{Eq:evolve}
\psi(\tau + \Delta \tau)&=
e^{-H \Delta \tau}  \psi(\tau)\\
&\approx e^{-\frac{1}{2} T \Delta \tau }
e^{-V \Delta \tau}e^{-\frac{1}{2} T \Delta \tau} \psi(\tau)
\notag
\end{align}
which is accurate to second order in $\Delta \tau$.  The order
parameter can then be evolved by taking successive Fourier (and
inverse Fourier) transforms of $\psi$ and multiplying by the factors
$e^{-\frac{1}{2}T \Delta \tau}$, $e^{-V \Delta \tau}$, and 
$e^{-T \frac{1}{2} \Delta \tau}$ respectively.  Such Fourier transforms account for the bulk
of the computational cost in this algorithm, thus using the efficient
Fast Fourier Transform algorithm is crucial.

A complication in the above occurs due to the nonlinearity of the GPE.
That is, $V$ in our above prescription for time evolution depends on
the density $n=|\psi|^2$, and it is at first unclear for what time
this quantity should be evaluated.  It is shown in \cite{javanainen06}
that, provided we use the most updated version of the time-dependent
density $n=|\psi|^2$, Eq.~(\ref{Eq:evolve}) will retain its second order
accuracy.
The final complication occurs from the rotational term in $H$ 
which is
\begin{equation}
R \equiv -\Omega L_z.
\end{equation}
We have neglected this term thus far since it is diagonalized in
neither position nor momentum space and therefore cannot be included
in either $T$ or $V$.  However, we note that $R$ commutes with both
$T$ and $V$ so we can write
\begin{align}
\label{Eq:evolve2}
\psi(\tau + \Delta \tau)&=
e^{-H \Delta \tau}  \psi(\tau)\\
&\approx e^{-\frac{1}{2} T \Delta \tau}
e^{-R\Delta \tau} e^{-V \Delta \tau}e^{-\frac{1}{2} T \Delta \tau} \psi(\tau).
\notag
\end{align}
Then we can perform a similar split-operator decomposition of
the additional term as
\begin{equation}
e^{-R\Delta \tau} \approx 
e^{\frac{1}{2} \hbar \Omega x p_y \Delta \tau}
e^{-\hbar \Omega y p_x \Delta \tau}e^{\frac{1}{2} \hbar \Omega x p_y \Delta \tau}.
\end{equation}
Evolution of $\psi$ by this factor can then be performed by taking
the partial Fourier transform of $\psi$, that is transforming over the $x$
variables but leaving the $y$ variables unchanged (or vice-versa).  
This completes the overview of the numerical method used to solve
the GPE.  As stated before, the generalization to the two-component
case is straightforward.  

\section{Results}
\label{Sec:results}

\begin{figure*}
\includegraphics[width=7.in]{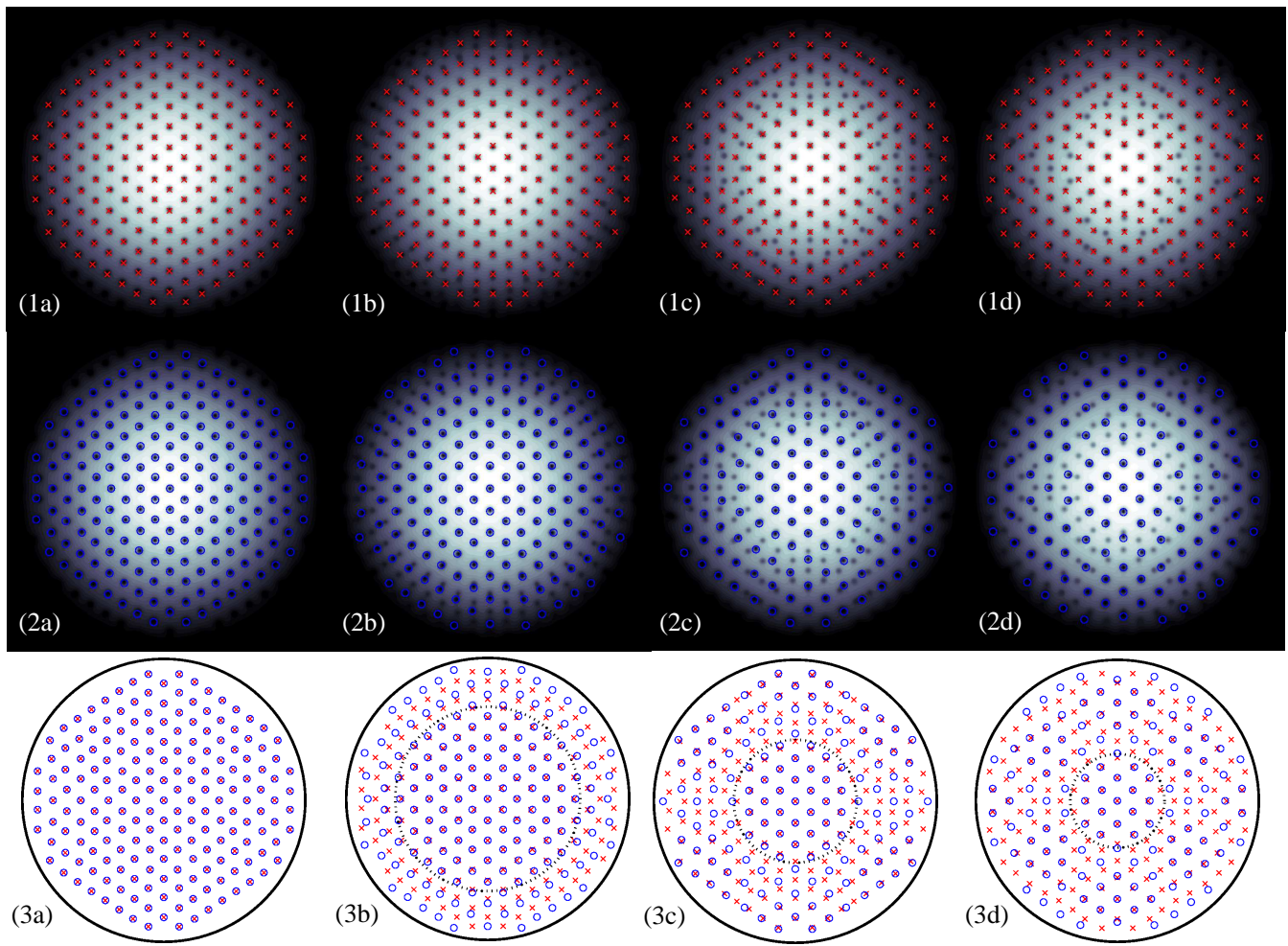}
\caption{Relaxed vortex lattices for several mass ratios.  Rows 1 and 2 
are the density profiles of species 1 and 2 respectively as a function of
position.  Superimposed over these images are the vortex positions marked
with red x's and black o's.  Row 3 gives the positions of both vortex 
species for comparison.  In this row, the dotted circle is the estimate
for the locking radius based on the two-body calculation showing good
agreement with the numerics.  Columns a, b, c, and d are for mass ratios
of $m_1/m_2 =$ 1.0, 1.2, 1.4, and 1.6 respectively. }
\label{Fig:main}
\end{figure*}

Next, we discuss the results of the numerical simulation. We first
provide the parameters which were used for the computations.  For
simplicity, we restrict our attention to the simplest case where
$g_1=g_2$, and we fix the interspecies scattering strength such that
$|g_{12}|/g_1=2/3$.  Furthermore, we take the number of particles in
each species to be the same: $N_1=N_2$.  We set the dimensionless
parameter defined as $\tilde{g}\equiv \frac{m}{\hbar^2} g_1 N_1$ to be
$\tilde{g}=2 \times 10^4$ which is in line with values from typical
experiments 
\footnote{Such a value would be obtained, for instance, for a condensate
of $3\times 10^6$ $^{87}$Rb atoms with condensate thickness 
$d_z=10 \mu m$.  This then gives $\tilde{g}= \frac{4\pi a}{d_z} N \approx
2 \times 10^4 $ where we take $a=106 a_0$ for
the $s$-wave scattering length.}.  
We take the two trapping potentials to be harmonic and
adjust their curvatures $\omega_1, \omega_2$ so that the density
profiles of the two species have the same Thomas-Fermi profiles.
Finally we rotate the system at 0.9 times the critical rate at which
the condensate becomes unstable
due to centrifugal forces.  We discretize the system
on a $200 \times 200$ grid, and propagate the system in 
imaginary time
intervals of $\Delta \tau= 0.01 \frac{1}{\hbar \omega_x}$.

We first consider the simplest case where the masses of the
two species are the same.  For this state we take the initial wavefunction
to be a perfect triangular lattice of vortices with density given by the
Feynman relation, Eq.~(\ref{Eq:Feyn}).  This structure is then relaxed
by evolving the wavefuntions in imaginary time using the methods
described in Sec.~\ref{Sec:numericalmethods}.  As expected these
vortex lattices remain fully locked.  The relaxed structures show
small deviations from the perfect triangular initial structure due
to the effects of the trap \cite{sheehy04}.  
The density profiles of these are shown
Fig.~\ref{Fig:main} in panels (1a) and (1b).  The positions of the
vortices are determined by analyzing the phases of the relaxed
wave functions.  
Using this relaxed structure as the initial state, we change the mass
ratios and propagate the wavefunctions in imaginary time until convergence.
Specifically, we consider the ratios of $m_1/m_2$ = 1.2, 1.4. and 1.6
as shown in Fig.~\ref{Fig:main}.  

To compare these numerical results to our estimate for the locking
radius described in Sec.~\ref{Sec:two-body}, we need to tailor
Eq.~(\ref{Eq:rc}) to the case of a harmonic trap.  We take the density
profiles used in Eq.~(\ref{Eq:rc}) to have the Thomas-Fermi form:
\begin{equation}
n_{1,2}=n_0 \left(1-\left(\frac{r}{R_{TF}}\right)^2 \right)
\end{equation}
where $n_0$ is the density at the center of the trap and 
$R_{TF}$ is the Thomas-Fermi radius
(note that we are only considering the case when the two 
condensates have the same radius).  Inserting this profile into 
Eq.~(\ref{Eq:rc}) (taking the correct dependence of the coherence
lengths on the density) one finds
\begin{equation}
r_c =r_c^0\sqrt{1-\left(\frac{r_c}{R_{TF}}\right)^2}
\end{equation}
where $r_c^0$ is Eq.~(\ref{Eq:rc}) evaluated for parameters at the
center of the trap.  This equation can then be 
solved for $r_c$ to obtain the renormalized value of the locking radius
\begin{equation}
\label{Eq:rc_ren}
r_c =\frac{r_c^0}{\sqrt{1+\left(\frac{r_c^0}{R_{TF}}\right)^2}}.
\end{equation}
This indicates that near the center of the condensate 
we will have $r_c \approx r_c^0$ as expected.  Also,  
the locking radius will never exceed the 
radius of the condensate as expected.  The reduction of the 
bare value of the locking radius can be qualitatively understood
as follows.  The Magnus force is proportional to the
superfluid density while the restoring force is proportional to this 
density squared.  Therefore near the edge of the condensate where
the density is considerably smaller than its value at the center,
the Magnus force will be favored thereby suppressing the locking
radius.

Shown in the third column of Fig.~\ref{Fig:main} are the positions of
the vortices of the two species, labeled with x's and o's.  The width
of these labels are roughly the size of the coherence length of the
condensates.  Superimposed on this is the locking radius predicted by
Eq.~(\ref{Eq:rc_ren}) (using Eq.~(\ref{Eq:rc_ren}) for the bar locking
radius) shown as a dotted line.  This shows that the analytic results
provide an excellent estimate of the locking radius.  Note that
due to the strong interactions between the two condensates, an unbound
vortex in one species will create a local minimum in the other.  Such
features can be seen in columns c and d of Fig.~\ref{Fig:main}.  These
local depletions should not be mistaken for vortices which are defined
by the phase behavior of the wavefunctions.

\section{Experimental Considerations and Concluding Remarks}
\label{Sec:exp}

The main requirement for realizing the locked state is a BEC composed
of a binary mixture of atoms having different masses and a negative
scattering length.  Such a transition could be tuned with an
interspecies Feshbach resonance which have been found in Li-Na
\cite{stan04} and Rb-K \cite{inouye04} mixtures.  Such mixtures have
respective mass ratios of 3.3 and 2.2.  Another promising experimental
system are mixtures of two isotopes of a particular atom.  For
instance, the interspecies scattering lengths of different species of
Yb have been analyzed in \cite{kitagawa08} and are often found to be
negative.  Since the mass ratios for different isotopes are closer to
unity, having a strong attractive interaction (often requiring a
Feshbach resonance) is unnecessary to reach the vortex locked state
for this case.

We also note that these results are closely related to the experiment
described in \cite{tung06}.  Here a single-component BEC is stirred by
a rotating optical lattice which acts as vortex pinning sites.  When
the optical lattice is rotated at the speed for which the density of
the pinning sites matches the density of the vortex lattice 
predicted by Eq.~(\ref{Eq:Feyn}), a completely locked state is observed.
Away from this resonance, a similar analysis to the above will predict
a disk of bound vortices.

In conclusion, we report the confirmation of the putative vortex
locked state proposed in \cite{barnett08}.  For this state, the two
superfluids and the stirring potential all rotate at different rates,
exhibiting an unusual effect due to the quantum mechanical nature of
superfluids.  In this paper, we showed that such a state exists within
a disk centered on the axis of rotation and whose size agrees well
with an analytic estimate.  Note that our numerical analysis did not
assume anything about the vortex-vortex attraction (unlike our
theoretical analysis, which assumes Eq. (\ref{Eq:E12}), and evolves
the Gross-Pitaevskii equations directly).  Our results (both
analytical and numerical) rely on approaching this state from the
fully locked state.  Experimentally, this is probably most easily
realized by controllably adjusting an interspecies Feshbach resonance.
Alternatively, one can use an optical lattice to control the effective
masses of the atoms by varying the lattice depth.

\acknowledgements
We would like to thank M. Porter and H.-P. B\"uchler for collaborations
on related previous work.  We would also like to thank L.  
Baksmaty for valuable advice on numerical methods.
This work was supported by the Sherman
Fairchild Foundation (RB); the Caltech SURF program (EC); 
and the Packard and Sloan Foundations, the
Institute for Quantum Information under NSF grants PHY-0456720 and
PHY-0803371, and The Research Corporation Cottrell Scholars program
(GR).

\appendix*

\section{Density profile for a single vortex}
\label{Sec:A1}

\begin{figure}
\includegraphics[width=3.4in]{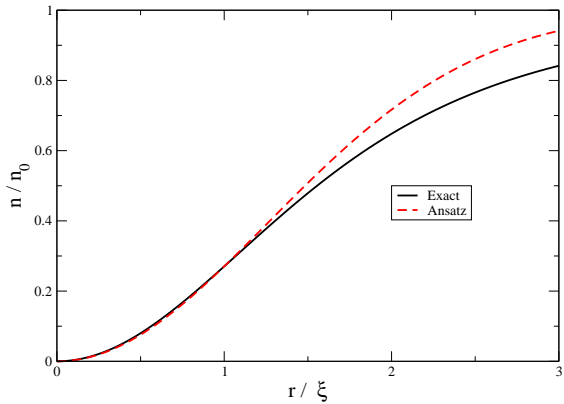}
\caption{Solid line: the density profile for a single vortex at $r=0$
  found from numerically solving Eq.~(\ref{Eq:GPEvort}).  Dashed line:
  ansatz density profile $n(r)=n_0(1-e^{-r^2/\lambda^2})$ where
  $\lambda$ is picked so that the two densities agree at one coherence
  length away from the vortex center.}
\label{Fig:vortprofile}
\end{figure}

In order to find the short-ranged interspecies vortex interaction, we
need to know the behavior of the density of the condensate about a
vortex.  To this end we consider the Gross-Pitaevskii equation for a
single component BEC having a vortex at the origin.  That is, we write
$\psi=f e^{i\theta}$ and take $\theta=\varphi$ where $\varphi$ is the
azimuthal angle from polar coordinates.  Substituting this into the
GPE leads to the following equation dictating the density profile
\begin{equation}
-\frac{\hbar^2}{2m} \frac{1}{r} \partial_r (r \partial_r f) + \frac{\hbar^2}{2m} \frac{f}{r^2} + g f^3
=\mu f.
\label{Eq:GPEvort}
\end{equation}
The density $n=f^2$ resulting from the numerical solution of this
equation is shown in Fig.~\ref{Fig:vortprofile}.

The numerical solution shows that the density behavior close to the
vortex core ($r \ll \xi$) is $n(r) \approx 0.340 \; n_0
\left(\frac{r}{\xi}\right)^2 $.  On the other hand, the far distance
behavior is found to be $n(r)=n_0 \left(1-
\left(\frac{r}{\xi}\right)^2\right)$.  To make our
work amenable to analytic treatment, we take the following ansatz for
the vortex profile:
\begin{equation}
n(r) = n_0 (1-e^{-r^2/\lambda^2})
\end{equation}
where $\lambda$ is a parameter on the order of the coherence length.
Note that while this ansatz has the correct form close to the vortex
core, the long distance behavior differs considerably.  Fortunately
our problem of vortex locking is dominated by the short-distance
behavior, and we choose $\lambda$ so that the two densities (numerical
and ansatz) agree at $r=\xi$ which requires $\lambda=1.781 \xi$, as
shown in Fig.~\ref{Fig:vortprofile}. 

Our positive results, confirming the vortex-locked state, also confirm
our intuition that the origin of the phenomena is in the short-ranged
attraction between vortices. As explained in
Ref. \onlinecite{barnett08} (but without proof), the algebraic decay
of the superfluid order parameter of a single votex does not imply
that vortices of one species, when in a lattice, exhibit a power-law
decaying force on the vortices on the other species. Unlike the
single-species vortex-vortex force, which is the result of the
inductive (kinetic) energy term in the Gross-Pitaevskii equations, the
interspecies force is a result of a density-density interaction. The
density suppression due to a single vortex occurs since the superflow
of the vortex effectively increases the mass terms $V_1,\,V_2$ in
Eq. (\ref{GGPP}). But in a lattice of vortices, the combined superflow
vector is nearly zero (i.e., negligible compared to $\hbar
/m_{\alpha}\xi_{\alpha}$), and, therefore, so is the respective
density suppression.


\end{document}